\documentclass[a4paper,lnbip]{svmultln}
\usepackage{graphicx}
\usepackage{latexsym}
\usepackage{float}
\usepackage{booktabs}
\usepackage{wrapfig}
\usepackage{todonotes}

\graphicspath{{img/}}

\begin{document}
\mainmatter 

\title{Low--Cost Eye--Trackers: Useful for Information Systems
Research?\thanks{This research is supported by Austrian Science Fund
(FWF): P26140--N15, P23699--N23. The final publication is available at Springer
via http://dx.doi.org/10.1007/978-3-319-07869-4\_14}}
\titlerunning{}
\author{Stefan Zugal \and Jakob Pinggera}

\institute{University of Innsbruck, Austria\\
\email{{stefan.zugal, jakob.pinggera}@uibk.ac.at}}
\maketitle

\begin{abstract}
Research investigating cognitive aspects of information systems is often
dependent on detail--rich data. Eye--trackers promise to provide respective
data, but the associated costs are often beyond the researchers' budget.
Recently, eye--trackers have entered the market that promise eye--tracking
 support at a reasonable price. In this work, we explore whether such
 eye--trackers are of use for information systems research and explore the
 accuracy of a low--cost eye--tracker (Gazepoint GP3) in an empirical
 study. The results show that Gazepoint GP3  is well suited for respective
 research, given that experimental material
  acknowledges the limits of the eye--tracker. To foster replication and
 comparison of results, all data, experimental material as well as the source code developed for this study are
 made available online.
\keywords{Eye--tracking, eye movement analysis, accuracy of fixations}
\end{abstract}

\setcounter{footnote}{0}
\section{Introduction}
\label{sec:introduction}
To facilitate the development of information systems, numerous modeling
languages, --methods and --tools have been devised over the last
decades~\cite{Mylo98}. Thereby, researchers found that not only the technical
perspective---such as correctness and expressiveness---are central requirements,
but also the human perspective needs to be taken into account. For instance, in
the field of business process management, researchers found that a good
understanding of a process model has a measurable impact on the success of a
modeling initiative~\cite{KVD+09}. Likewise, business process vendors and
practitioners ranked the usage of process models for understanding business
processes as a core benefit~\cite{Ind+09}.

To support humans in their interaction with artifacts created during the
development of information systems, e.g., models or source code, various
research methods have been followed. For instance, researchers analyzed
communication protocols gathered in modeling workshops~\cite{Hada13}, sought to
adapt theories from cognitive psychology~\cite{Zuga13}, investigated think aloud
protocols~\cite{Hai+13} or adopted techniques from eye movement analysis for
assessing the comprehension of business process models~\cite{PeMe13}. In this
work, we focus on the role of eye movement analysis, as we think that the
adoption of eye movement analysis is still below its full potential. In
particular, it seems plausible that the  costs of eye tracking
infrastructure poses a considerable burden for the adoption of eye movement
analysis~\cite{HeDu10}.\footnote{High--precision eye--trackers can cost more
than several ten--thousand US\$, see:\\
\url{http://www.arringtonresearch.com/prices.html} (accessed February 2014).} To
counteract this problem, efforts have been undertaken for developing
 eye--trackers at a low price by assembling off--the--shelf components,
 e.g.,~\cite{DoBD06}. However, it is questionable in how far researchers who
 are not deeply involved in the peculiarities of assembling hardware are able to
 set up such an infrastructure on their own. Rather, we see a big potential in
 low--cost ready--to--use eye--trackers that have entered the market, seeking to
 compete with the high--priced versions.\footnote{For instance:
 \url{http://theeyetribe.com/}, \url{http://gazept.com/},\\
 \url{http://mygaze.com/} (accessed February 2014)} Particularly in
 times of shortened research budgets, respective cost--efficient infrastructure
 seems indispensable.

In this sense, the research question investigated in this study can be defined,
as follows: \textit{Are low--cost eye--trackers useful for information
systems research? If yes, which limitations apply?} To approach this research question,
we bought \textit{Gazepoint GP3}\footnote{\url{http://gazept.com/products/}
(accessed February 2014)} and employed it in an empirical study for assessing
its accuracy. Likewise, the contribution of this
work is threefold: First, we report on the accuracy of Gazepoint GP3
with respect to the detection of fixations. Second, we use respective data for describing how
experimental material, e.g., models, source code or tools, should be designed so
that acceptable error rates can be expected. Third, we provide the source code
used for this study, thereby providing an infrastructure for evaluating Gazepoint GP3 in different settings. Likewise, the remainder of this paper is
structured, as follows.
Section~\ref{sec:background} introduces background information on
eye--tracking.
Then, Section~\ref{sec:experimental_design} describes the experimental design of
this study, whereas results are described in Section~\ref{sec:results_and_discussion}. Finally,
Section~\ref{sec:related_work} discusses related work and
Section~\ref{sec:conclusion} concludes  with a summary and an
outlook.

\section{Eye Movement Analysis}
\label{sec:background}
Before describing the experimental design followed in this study, we briefly
introduce basic concepts related to eye movement analysis (for a more detailed
introduction, see e.g.,~\cite{Duch07}). The fundamental idea of eye movement
analysis is capturing the position a person is currently focusing on. To this
end, usually the \textit{pupil center corneal reflection method}~\cite{Ohn+02} is
adopted, in which the center of the pupil is computed by assessing the corneal
reflection (Purkinje reflection) through infrared light. Thereby, either remote
systems (i.e., video and infrared cameras that are affixed to a table) or
head--mounted systems (i.e., devices that are fixed on the person's head) are
employed~\cite{Rau+12}. However, only capturing the position a person is looking
at is not enough, as it is known that high--resolution visual information
input can only occur during so--called \textit{fixations}, i.e., when the person
fixates the area of interest on the fovea, the central point of highest
visual acuity~\cite{Posn95}. These fixations can be detected when the velocity
of eye movements is below a certain threshold for a pre-defined duration~\cite{JaKa03}.
Using eye fixations, we can identify areas on the screen the person is focusing
attention on~\cite{FuSa08}, e.g., features of the modeling environment or
modeling constructs. Due to the central role of fixation for processing visual
information, we focus on the accuracy of fixation detection in the following.
\section{Experimental Design}
\label{sec:experimental_design}
The goal of this empirical investigation is to determine whether low--cost
eye--trackers provide enough accuracy to be of use for
information systems research. As discussed in Section~\ref{sec:background}, fixations are of
central interest, hence next we describe the experimental design followed for
investigating the accuracy of fixations.

\paragraph{Experimental Procedure} 
The procedure followed in this experimental design consists of 5 steps. First,
the subject is informed about potential risks involved in participating in the
experiment and that all data is collected anonymously. Second, the eye tracker
is calibrated by a 9--point calibration, as provided by the Gazepoint GP3 API.
Third, the first visual task is presented to the subject, which basically asks
the subject to look at specific points at the screen (details are provided in
Paragraph \textit{Visual Tasks}). Fourth, as it may be the case that the subject
was not entirely focused on the visual task, the task is repeated once more.
Finally, each experimental session is concluded by administering a survey about
demographical information. To enable replication, the entire experimental
material and data is freely available.\footnote{The experimental material and
data are available at:\\
\url{http://bpm.q-e.at/eye-tracking-accuracy}}

\begin{figure}[htb]
  \centering
  \includegraphics[width=\textwidth]{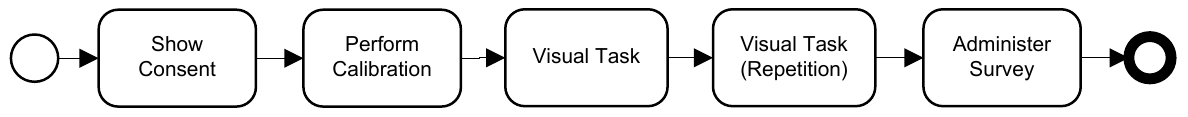}
  \caption{Experimental design}
  \label{fig:experimental_design}
\end{figure}

\paragraph{Visual Tasks}
The visual tasks administered in this study were designed for measuring
accuracy, i.e., computing the difference between the position on the screen the
subject looked and the fixations the eye tracker computed. To this end, subjects
are asked to look at a specific position for a given time interval. To ensure
that subjects were indeed looking at this specific point, we asked to press a
key as soon as the subject fixated on the point. As it is then known where the
subject looked, fixations can be compared with this position. Technically, we
implemented a Java component, which displays a configurable list of points
according to the following procedure:
\begin{enumerate}
  \item Fill the screen with white color
  \item For each point in the configured lists of points
  	\begin{enumerate}
  		\item Draw solid black point on the screen (10 * 10 pixel) at given
  		position
  		\item Wait until user presses arbitrary key
  		\item Capture the moment when the user presses the key
  		\item Wait for 500 ms
  		\item Fill the screen with white color
	\end{enumerate}
\end{enumerate}

Using this mechanism, we configured 9 points, equally
distributed on a grid on the screen. Since the employed eye--tracker is not
able to track points outside the screen, we avoided points near the
ending of the screen, i.e., the grid started at point (0.25 * width,
0.25 * height) and ended at point (0.75 * width, 0.75 * height). Apparently,
subjects require time to locate and fixate on the current point. Hence, we only
captured data between the moment when the subject pressed the key and
the next 500 ms. The duration in which
fixations are collected for analysis, i.e., 500 ms, is a
trade--off between the amount of data points that can be collected and the
quality of the data. In a longer time window more data can be collected,
but at the same time it becomes likelier that the subject gets distracted, and
vice versa. The eye--tracker's cameras operate at 60 Hz, likewise 60 points
are recorded per second. Thus, 500 ms, resulting in approximately 30 data points
per subject seem to be an acceptable trade--off.

\paragraph{Subjects}
The population under examination in this study were all persons that may
participate in eye--tracking research. The tasks involved in this experimental
design does not require special training, rather basic reading skills are
sufficient. However, during the preparation of the software displaying the
visual tasks, we observed that the eye--tracker could not properly handle the
reflections of glasses (no complications could be observed for persons wearing
contact lenses). Hence, we exclude persons wearing glasses from our experimental
setup.

\paragraph{Experimental Setup}
For performing the eye movement analysis, a table mounted eye tracker, i.e.,
Gazepoint GP3, was used, recording eye movements at a frequency of 60 Hz. The
visual tasks were performed on a 20'' monitor with a resolution of 1600 * 1200
pixels and a dimension of 40 cm * 30 cm. In addition, we attached a second
monitor, on which the eye--tracking software was running, allowing to monitor
whether subjects were within the area accessible to the eye--tracker's cameras.
The second monitor was positioned away from the main monitor, allowing the
subject to fully concentrate on the visual tasks. The subject was seated
comfortably in front of the screen in a distance of approximately 65cm (as
recommended by the eye--tracker's manual). To minimize undesired
fluctuations regarding light, we closed blinds of the office windows.

\paragraph{Response Variable}
The interest of this study is to examine the accuracy of an eye--tracker with
respect to the detection of fixations. Hence, the response variable of this
study is the distance between the point the subject was supposed to look at and
the corresponding fixations measured by the eye--tracker, subsequently referred
to as \textit{error}. As described in Paragraph \textit{Visual Tasks}, we stored
the moment when the subject pressed any key and showed the point for
another 500 ms; only data points collected during this time window are used for
analysis. For all of the data points falling into this time frame, in turn, the
error is computed as the Euclidean distance between the measured fixation and
the point displayed.
Furthermore, data points are only taken into account when considered to be a
valid fixation according to the eye--tracker's internal fixation filter.

\paragraph{Instrumentation and Data Collection} 
To allow for an efficient collection and analysis of data, we implemented the
experimental procedure shown in \figurename~\ref{fig:experimental_design} as an
experimental workflow in Cheetah Experimental Platform (CEP)~\cite{PiZW10}. In
other words, each activity from the experimental procedure was supported by a
Java component, which in turn was executed in the order prescribed by the
experimental procedure. Thereby, CEP provided ready--to--use components for
displaying a consent dialog and administering a survey, whereas eye--tracking
related components had to be implemented.

\newcommand{\rHead}[1]{\multicolumn{1}{p{1.5cm}}{\hfill\textbf{#1}}}
\newcommand{\lHead}[1]{\multicolumn{1}{p{1.6cm}}{\textbf{#1}}}
\section{Results and Discussion}
\label{sec:results_and_discussion}
So far we described the experimental design adopted in this study. In the
following, we focus on the execution of the empirical study in
Section~\ref{sec:results}, discuss implications in
Section~\ref{sec:discussion}, and present its limitations in
Section~\ref{sec:limitations}. We would like to repeat at this point that all
data collected is available on--line.\footnote{All data collected in this
study is available at:\\
\url{http://bpm.q-e.at/eye-tracking-accuracy}}

\subsection{Experimental Execution}
\label{sec:results}
In the following, we describe the preparation of the experiment, before we turn
to the execution of the experiment and subsequently present the collected
data.

\paragraph{Experimental Preparation}
Preparation for this empirical study included acquiring the eye--tracker,
implementing components accessing the eye--tracker's API, configuring the
experimental procedure in CEP and acquiring subjects. Since our
experimental procedure does not involve any particular skills besides reading,
we relied on a convenience sample, i.e., we acquired friends and co--workers at the
Department of Computer Science at the University of Innsbruck. As described in
Section~\ref{sec:experimental_design}, none of the persons wore glasses during
the experiment. However, we included subjects that wore glasses in daily life,
but were able to read texts without glasses (these subjects participated
without glasses).

\paragraph{Experimental Execution}
The eye--tracking sessions were performed in February 2014 at the University of
Innsbruck, where 16 subjects participated. However, for one subject the
eye--tracker had problems identifying the subject's pupils, so we decided to
exclude the data from analysis, leaving 15 data sets for analysis. Due to the
nature of eye--tracking, only one subject could participate at a time. In this
way, each subject could be welcomed, introduced to the experimental procedure
and guided through the eye--tracking session. To reward and motivate subjects, a
plot showing all fixations was produced immediately \textit{after} each session,
allowing subjects to see how well they performed.

\paragraph{Data Validation}
To assess whether the collected data is valid, we plotted the results of each
visual task for each subject and inspected the plots for abnormalities. Interestingly,
the analysis revealed that for certain subjects the fixations of the visual task
that was intended as familiarization were more accurate than for the second
visual task. Knowing that subjects usually need a little training to get acquainted
with tasks, these results seemed implausible. However, a discussion revealed
that certain subjects had remembered where the next dot would appear and
did not fully concentrate on the current point, but already moved on to the next
point. To compensate for this shortcoming, we compared the results for the first
and the second visual task and selected the task showing the \textit{lower
median} of error values. We argue that this procedure is acceptable,
since it can be assumed that the eye--tracker's accuracy is stable for
the \textit{same} task and \textit{same} subject. In other words, fluctuations
can be rather attributed to subject--related factors, e.g., familiarization with
the task or the discussed anticipation of points, hence selecting the visual
task with lowest errors will probably result in selecting the visual task with
the least influence of subject--related factors.

\paragraph{Results}
Next, we describe the data obtained in this study from three perspectives.
First, we look into demographical statistics, second, turn to error quantiles
and, third, discuss the results of one subject. A summary of demographical data
can be found in Table~\ref{tab:demographics}. Subjects were on average 30.67
years old ($SD = 3.27$) and 33.3\% female. None of the subjects reported 
eye diseases or wore glasses. However, 5 subjects used
contact lenses during the experiment. 

\begin{table}
\centering
\begin{tabular}{p{5.5cm}l}
\toprule
\lHead{Variable}& \textbf{Data}\\ \midrule
Age & Min: 26, Max: 41, M: 30.67, SD: 3.27\\
Gender & Female: 5 (33.3\%), Male: 10 (66.7\%) \\
Eye diseases & 0 (0\%)\\
Glasses during experiment & 0 (0\%)\\
Contact lenses during experiment & 5 (33.3\%)\\
\bottomrule
\end{tabular}
\caption{Demographical data}
\label{tab:demographics}
\end{table}

To give an overview of the collected fixations, we have summarized the
error occurred during calibration ($\varepsilon_{calib}$), quantiles describing
the error distributions ($Q_{95}$ to $Q_{80}$) as well as the median of errors.
In particular, as shown in Table~\ref{tab:fixations_pixel}, the average error
measured during calibration was 45.20 pixel and quantiles range from 116.38
($Q_{95}$) to 58.00 ($Q_{80}$); the median of errors was 32.20 pixel. All in
all, 4,122 fixations were captured, of which 3,869 were considered to be
valid according to the eye--tracker's fixation filter. Furthermore, it can be
observed that fixations seem to be rather homogeneous---a box plot of median
values only detected $S_{8}$ and $S_{13}$ as outliers.

\begin{table}[htb]
\centering
\begin{tabular}{lrrrrrrrr}
\toprule
\lHead{Subject} & \rHead{\boldmath{$\varepsilon_{calib}$}} &
\rHead{\boldmath{$Q_{95}$}}& \rHead{\boldmath{$Q_{90}$}}& \rHead{\boldmath{$Q_{85}$}}&
\rHead{\boldmath{$Q_{80}$}} & \rHead{\textbf{Median}}\\
\midrule
$S_{1}$ & 47.14 & 59.91 & 47.30 & 37.64 & 31.32 & 21.00\\
$S_{2}$ & 62.59 & 66.22 & 62.77 & 59.48 & 50.45 & 30.27\\
$S_{3}$ & 53.19 & 92.46 & 87.86 & 68.18 & 65.37 & 44.91\\
$S_{4}$ & 28.86 & 46.10 & 37.95 & 34.79 & 32.70 & 17.03\\
$S_{5}$ & 43.96 & 103.62 & 97.51 & 58.18 & 52.35 & 27.29\\
$S_{6}$ & 49.67 & 43.42 & 38.12 & 31.40 & 29.83 & 19.72\\
$S_{7}$ & 38.24 & 96.13 & 73.82 & 71.69 & 70.26 & 27.78\\
$S_{8}$ & 74.56 & 180.50 & 166.21 & 146.01 & 128.23 & 56.63\\
$S_{9}$ & 37.74 & 86.76 & 84.91 & 51.43 & 48.26 & 34.07\\
$S_{10}$ & 47.74 & 86.59 & 66.94 & 61.55 & 57.01 & 34.46\\
$S_{11}$ & 43.23 & 121.25 & 117.69 & 67.78 & 66.48 & 33.94\\
$S_{12}$ & 43.48 & 88.20 & 76.55 & 71.87 & 65.51 & 33.62\\
$S_{13}$ & 42.85 & 181.11 & 178.02 & 173.00 & 152.27 & 116.87\\
$S_{14}$ & 24.02 & 51.88 & 50.25 & 48.26 & 47.10 & 35.18\\
$S_{15}$ & 40.65 & 43.57 & 41.76 & 40.31 & 38.28 & 21.93\\
\midrule
Total & 45.20 & 116.38 &  84.96 & 66.73 & 58.00 & 32.20\\
\bottomrule
\end{tabular}
\caption{Results for fixations (in pixel)}
\label{tab:fixations_pixel}
\end{table}

So far, we have discussed the error distributions measured in pixel. To give an
impression what these errors mean with respect to screen size, we have listed
the error in mm in Table~\ref{tab:fixations_mm}. As described in
Section~\ref{sec:experimental_design}, the eye--tracking sessions were performed
on a screen with an extent of 1600 * 1200 pixel and a screen size of 40 * 30
cm, i.e., 4 pixels were displayed per mm. In other words, values listed in
Table~\ref{tab:fixations_mm} were computed by dividing the pixel--values from
Table~\ref{tab:fixations_pixel} by factor 4.

\begin{table}[htb]
\centering
\begin{tabular}{lrrrrrrrr}
\toprule
\lHead{Subject} & \rHead{\boldmath{$\varepsilon_{calib}$}} &
\rHead{\boldmath{$Q_{95}$}}& \rHead{\boldmath{$Q_{90}$}}& \rHead{\boldmath{$Q_{85}$}}&
\rHead{\boldmath{$Q_{80}$}} & \rHead{\textbf{Median}}\\
\midrule
$S_{1}$ & 11.78 & 14.98 & 11.82 & 9.41 & 7.83 & 5.25\\
$S_{2}$ & 15.65 & 16.55 & 15.69 & 14.87 & 12.61 & 7.57\\
$S_{3}$ & 13.30 & 23.11 & 21.97 & 17.05 & 16.34 & 11.23\\
$S_{4}$ & 7.22 & 11.52 & 9.49 & 8.70 & 8.17 & 4.26\\
$S_{5}$ & 10.99 & 25.90 & 24.38 & 14.55 & 13.09 & 6.82\\
$S_{6}$ & 12.42 & 10.85 & 9.53 & 7.85 & 7.46 & 4.93\\
$S_{7}$ & 9.56 & 24.03 & 18.46 & 17.92 & 17.57 & 6.95\\
$S_{8}$ & 18.64 & 45.12 & 41.55 & 36.50 & 32.06 & 14.16\\
$S_{9}$ & 9.44 & 21.69 & 21.23 & 12.86 & 12.06 & 8.52\\
$S_{10}$ & 11.94 & 21.65 & 16.74 & 15.39 & 14.25 & 8.62\\
$S_{11}$ & 10.81 & 30.31 & 29.42 & 16.94 & 16.62 & 8.49\\
$S_{12}$ & 10.87 & 22.05 & 19.14 & 17.97 & 16.38 & 8.40\\
$S_{13}$ & 10.71 & 45.28 & 44.50 & 43.25 & 38.07 & 29.22\\
$S_{14}$ & 6.00 & 12.97 & 12.56 & 12.06 & 11.77 & 8.79\\
$S_{15}$ & 10.16 & 10.89 & 10.44 & 10.08 & 9.57 & 5.48\\
\midrule
Total &  11.30 & 29.10 &  21.24 & 16.68 & 14.50 & 8.05\\
\bottomrule
\end{tabular}
\caption{Results for fixations (in mm)}
\label{tab:fixations_mm}
\end{table}

\begin{figure}[htb]
  \begin{center}
  \includegraphics[width=0.9\textwidth]{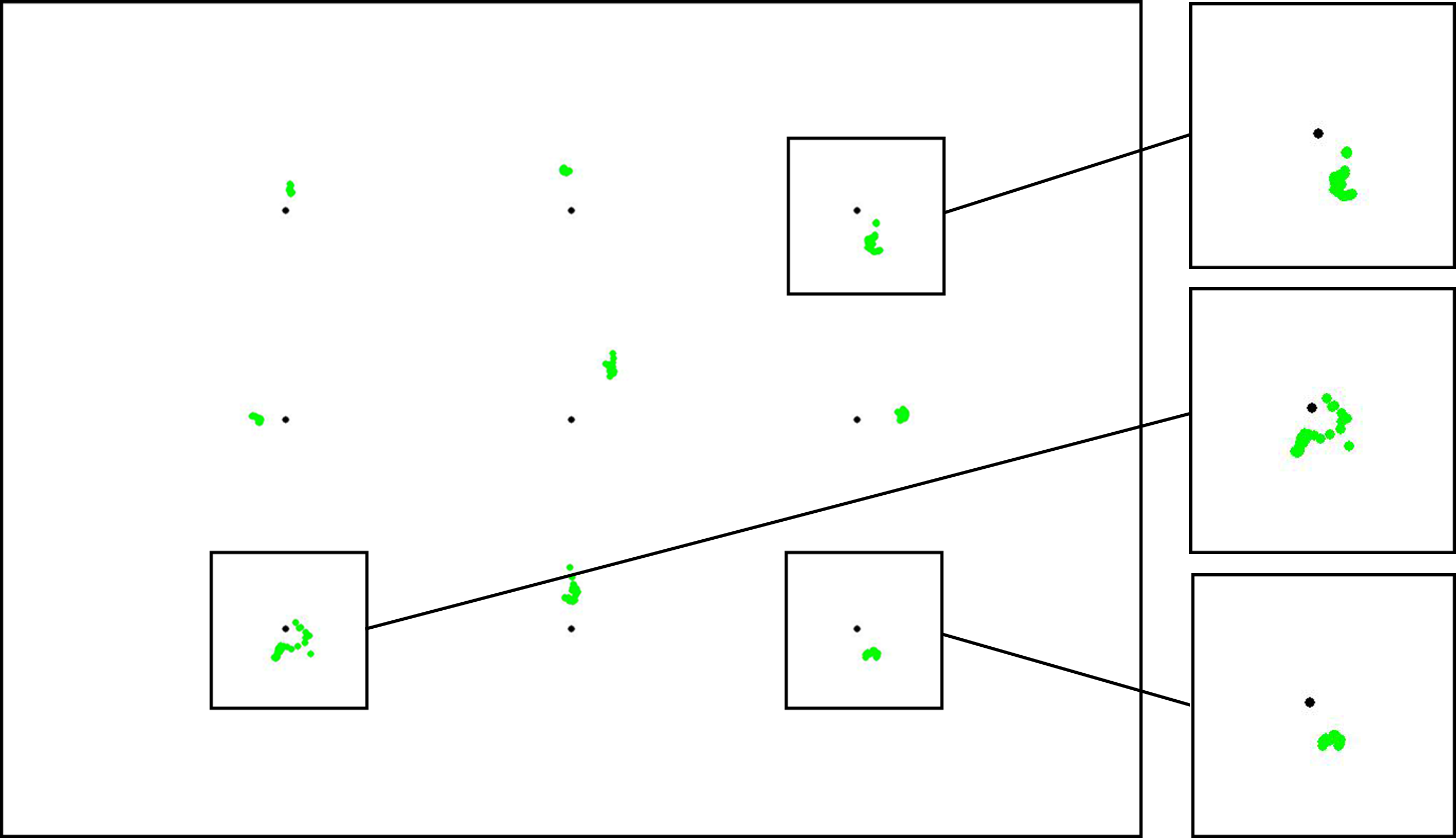}
  \end{center}
  \caption{Fixations measured for subject $S_3$}
  \label{fig:fixations}
\end{figure}

For visualizing how these ranges of errors relate to a screen of 1600 * 1200
pixel, we have selected the fixations for a subject with approximately average
errors, i.e., subject $S_3$, and visualized the results in
\figurename~\ref{fig:fixations}. In particular, the box to the left represents
the screen with the 9 points at which $S_3$ was asked to look at. The green
dots, in turn, represent the fixations as obtained through the eye--tracker. To
the right, we have selected three regions to show patterns we could observe in
the data. Mostly, as shown in the square at top right, fixations were measured
in a rather small region, which does not necessarily directly overlap with the
point the subject was supposedly looking. Also, as shown in the square in the
middle right, fixations may have also been scattered over a larger region. This
behavior could particularly be observed for subjects with large errors, e.g.,
subject $S_8$. Finally, as shown in the square bottom right, fixations for
almost the same location were reported. However, the fixations were not
necessarily directly at the location the subject was supposedly looking. We do
not want to speculate here about potential reasons for these results, but rather
give some visually accessible perspective on the data.

%
%

\subsection{Discussion}
\label{sec:discussion}
So far we described the data, next we discuss implications with respect to
the adoption of low--cost eye--trackers. Basically, certain criteria must be
fulfilled so that Gazepoint GP3 can be used in a
meaningful way.\footnote{We are not aware of any published
studies utilizing this particular device.} As described in
Section~\ref{sec:experimental_design}, we could not manage to get the eye--tracker working for subjects wearing glasses, since reflections of the glasses were confused with reflections of the eyeball.
Also, subjects with small eyes caused significant troubles in identifying
pupils. Interestingly, also particularly glossy hair caused problems---in fact,
for one subject it was only possible to conduct the session after the subject
covered the hair. In addition, direct sunlight complicated the identification of
fixations. However, as all of these problems could be identified and resolved
during calibration, they presumably did not influence the results of this study.
Regarding the accuracy promised by the vendor, the manual specifies an accuracy
of \textit{$0.5^{\circ}$} to \textit{$1^{\circ}$ of visual angle}. Assuming that
the line of sight, error and screen form a right angle, an error of $1^{\circ}$
results approximately in an error of 1.05 cm, or 42 pixels (tan(1) * 60 cm
$\approx$ 1.05 cm). These promises are in line with our findings: As shown in
Table~\ref{tab:fixations_mm}, the median error for fixations was 8.05 mm.

\setlength{\intextsep}{10pt}
\begin{wrapfigure}{r}{0.5\textwidth}
  \begin{center}
  \includegraphics[width=0.5\textwidth]{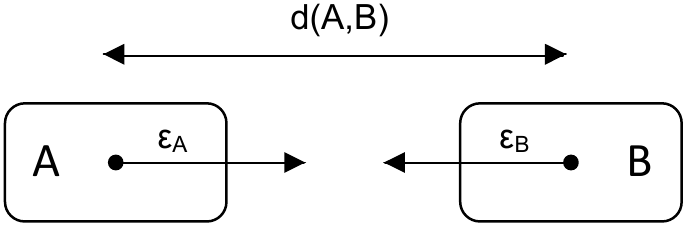}
  \end{center}
  \caption{Distinguishing objects}
  \label{fig:error}
\end{wrapfigure}

To finally answer the research question approached in this work, i.e., whether
low--cost eye--trackers are useful for information systems research, we discuss
whether these accuracies are good enough for identifying where subjects looked.
For this purpose, consider the illustration  in
\figurename~\ref{fig:error}, showing two objects (A and B) that should be
identified in a study, e.g., activities in business process models, source
code snippets or parts of a user interface.
 In particular, the figure shows objects A and B, the distance between the
 center of A and the center of B ($d(A, B)$) as well as the errors involved in
 measuring fixations for the position of A
($\varepsilon_A$) and the position of B ($\varepsilon_B$). Conservatively
assuming that errors are always directed toward the opposite object, fixations
can only be unambiguously assigned to an object if the distance between the
objects is smaller than the sum of error, i.e., $d (A, B)$ $<$ $\varepsilon_A +
\varepsilon_B$. As listed in Table~\ref{tab:fixations_pixel}, the median error
was 32.20 pixel, i.e., if the difference between 2 objects is less than 64.40,
the objects cannot be distinguished anymore. Likewise, when increasing
the distance between objects, the probability of properly identifying objects
increases. Whether these distances can be achieved in a meaningful way then
depends on the specific research question. On the one hand, for instance,
studies that investigate reading source code respective distances between source
code characters seems infeasible. On the other hand, for instance, when
evaluating, the user adoption of recommendations (e.g.,~\cite{Zan+11}),
respective distances in a user interface can easily be achieved. Hence,
depending on the specific research question, the accuracy of Gazepoint GP3 may
or may not be sufficient.

\vspace{-0.3cm}

\subsection{Limitations}
\label{sec:limitations}
As in every empirical study, the results have to be seen in the light of several
limitations. First, the question arises in how for results can be generalized to
the population under examination, i.e., all persons that potentially participate
in eye--tracking studies. Since only data of 15 subjects was used for analysis,
results need to be generalized with care. Similarly, subjects were of rather
young age, i.e., on average 30.67 years old, hence it is not clear whether
results also apply to older persons. Second, it must be acknowledged that the
performance of the visual tasks depends on whether subjects were indeed looking
at the points they were asked to focus on. For instance, subjects remarked that
they knew where the next point would appear, making it difficult to focus on the
current point. By selecting the task with higher accuracy, we sought to
compensate for this issue. Third, results are applicable only to Gazepoint GP3
and cannot be generalized to other eye--tracking devices as produced by other
vendors. Finally, we tried to provide similar settings for all subjects, e.g.,
same tasks, instructions and monitor size. However, we could not fully control
all external influences, such as light.
Also, we would like to emphasize that it is out of the scope of this
contribution to compare these results with high--end eye--trackers.

\vspace{-0.4cm}
\section{Related Work}
\label{sec:related_work}
In this work, we focused on eye--tracking in information systems research,
seeing eye--trackers from the perspective of users. However, also on the
developers' side vivid research activities can be observed, e.g., investigating
the feasibility of self--built eye--tracking systems~\cite{DoBD06}, developing
support for eye--tracking on mobile devices~\cite{Dyb+12} and designing new
algorithms for the detection of eye movements~\cite{ViBG12}. However, these
works rather focus on the development of new methods and applications than on
evaluation the feasibility, as done in this work. Regarding the use of
eye--tracking, applications in a variety of domains can be observed. For
instance, experiments have been conducted for investigating the
understandability of UML models~\cite{PoGu10} and the interpretation of data
models~\cite{NoCr99}. Similarly, a research agenda for investigating user
satisfaction has been proposed in~\cite{Hog+11}. Other works employed
eye--tracking for investigating process model comprehension~\cite{PeMe13} or for
inspecting the way how modelers create process models~\cite{PFM+13}. However,
all these works focus on the direct application of eye movement analysis rather
than seeking to examine its usefulness, as done in this work. Even though we
 focus on eye--tracking, it is clearly not the only promising approach
for investigating cognitive aspects in information systems. For instance, think
aloud protocols, i.e., the thoughts subjects uttered during an empirical study,
may be used to get insights into the cognitive processes involved in working
with information systems artifacts~\cite{Hai+13,SeMW13}. Also, researchers have
sought to transfer theoretical concepts from other domains, e.g., cognitive
psychology, for advancing information systems research~\cite{Zug+12b,Zug+11a}
and to conduct controlled experiments, e.g.,~\cite{PZW+10b,CVR+12}. It is
important to stress that these approaches should be not seen as competing.
Rather, best results can be expected by combining two or more paradigms, i.e.,
through method triangulation~\cite{Jick79}.
\vspace{-0.5cm}
\section{Summary and Conclusion}
\label{sec:conclusion}
\vspace{-0.1cm}
In this work, we set out to examine the accuracy of a low--cost
eye--tracker regarding the detection of fixations. In an empirical study, we
asked participants to look at specific positions at the screen and recorded the
computed fixations. The analysis of errors showed
that the median error lies well within the accuracy promised by the vendor. In a
next step, we discussed the implications of these findings with respect to the
development of experimental material, showing that the eye--tracker is well
suited---given that elements at the screen are placed in proper distance. Thus,
we conclude that if certain preconditions are fulfilled, e.g., not wearing
glasses, appropriate light and covering glossy objects, Gazepoint GP3
appears to be a suitable choice for affordable eye--tracking studies.

Particularly for research that focuses on cognitive aspects, multiple
perspectives as well as detail--rich data is indispensable.  By providing data
about the eye--tracker's accuracy, respective recommendations for developing
experimental material and making available the source code involved in
this study, we hope to help spreading and establishing eye--tracking for
information systems research in general and research on cognitive aspects in
particular. Regarding our research, we seek to employ the findings of this study
for developing algorithms that automatically detect the modeling element a
process modeler is looking at. With respective support, the cognitive processes
involved in creating process models could then be investigated in an even more
efficient and detailed manner.
\vspace{-0.5cm}

\bibliographystyle{splncs}
\bibliography{literature}

\end{document}